\documentclass[11pt]{article}

\usepackage{amsmath, epsfig, cite}

\usepackage{amssymb}

\usepackage{amsfonts}

\usepackage{latexsym}

\newtheorem{theorem}{Theorem}[section]

 \numberwithin{equation}{section}

 \makeatletter \@addtoreset{equation}{section} \makeatother

\title{An improved bound on the Maximum Agreement Subtree problem}

\author{
L\'aszl\'o  A. \ Sz\'ekely\footnote{This author was supported in
part by the NIH NIGMS contract  1 R01 GM078991-01, by the NSF DMS
contract 0701111, by a Marie
Curie Fellowship HUBI MTKD-CT-2006-042794, and by the 2007
Phylogeny program of the Isaac Newton Institute, Cambridge, where this work
started.},\\
Department of Mathematics,\\
University of South Carolina, {\tt szekely@math.sc.edu}\\
Mike Steel \footnote{This author
was supported in part by the New Zealand Marsden Fund and the Allan Wilson Centre for
 Molecular Ecology and Evolution.},\\
Biomathematics Research Centre, \\University of Canterbury, New Zealand,
 {\tt M.Steel@math.canterbury.ac.nz }\\
}

\begin{document}

\maketitle

\begin{abstract}
We improve the lower bound on the extremal version of the
Maximum Agreement Subtree problem. Namely
we prove that two binary trees on the same $n$ leaves
have subtrees with the same     $\geq c\log\log n$ leaves which are
homeomorphic, such that homeomorphism is identity on the leaves.
\end{abstract}

\newpage

\section{Introduction}

A {\it phylogenetic X-tree} is
a binary tree in which the leaves are labelled bijectively with
labels from a set $X$ (usually $ \{1,2,...,n\}$) and internal vertices
are unlabelled.
Two phylogenetic X-trees are considered
 the same, if there is a label-preserving
graph isomorphism between them.

If $ T$ is phylogenetic X-tree  and $ Y\subseteq X$  is a set of labels,
then the {\it  induced binary subtree }
$T|_Y$ is defined as follows:
(a) take the subtree induced by $Y$ in $T$, and (b)
substitute paths in which all internal vertices have degree 2
 by edges. $T|_Y$ is a phylogenetic Y-tree (see Fig. 1).

If $|Y|=4$, the induced binary subtree is often identified with an
unordered partition of $Y$ into two two-element sets, obtained by
removing the (unique) internal edge of  $T|_Y$. This partition
is known as {\it quartet split}. It has been known that the ${\binom n 4}$
quartet splits of phylogenetic X-tree with $|X|=n$ determine the
phylogenetic tree through a polynomial time algorithm. This was
first observed in 1981 by Colonius and Schultze \cite{colonius},
in the context of stemmatology, and was developed further in 1986 by
Bandelt and Dress \cite{ban86}.

An important algorithmic problem, known as the {\em Maximum Agreement Subtree
Problem}, is the following: given two phylogenetic $X$--trees, find a
common induced binary subtree of the largest possible size.

\begin{figure}[ht] \begin{center}
\resizebox{12.5cm}{!}{
\input{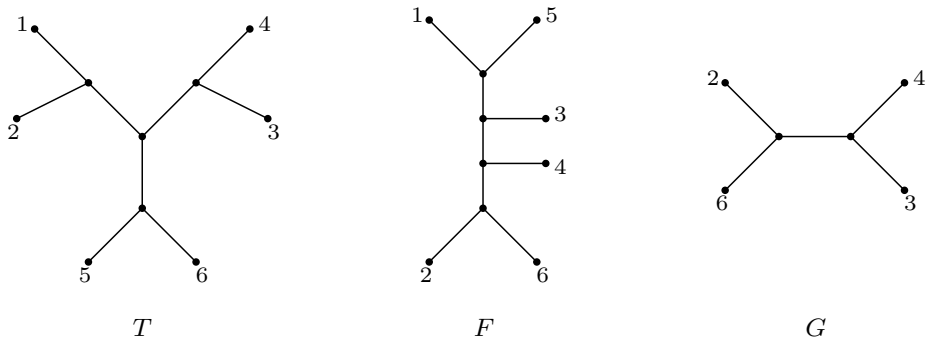}
}
\caption{For $X=\{1,2,3,4,5,6\}$ and the two phylogenetic $X$--trees shown ($T$ and $F$), a maximum agreement
subtree is the phylogenetic tree $G=T|_Y=F|_Y$ shown, where $Y=\{2,3,4,6\}$.}
\end{center}
\label{motiv}
\end{figure}

This problem has a history that spans more than 25 years, from papers in the early 1980s by Gordon
 \cite{gordon}, and
Finden and Gordon \cite{fgordon}; to its implementation in the late 1990s in the widely-used phylogenetic software
PAUP \cite{PAUP}.  Somewhat surprisingly, this
problem can be solved in polynomial time \cite{stewar} (see also \cite{GKKMcM} and \cite{KKMcM95}).

Here we focus on the extremal version of the problem. Let
${\rm mast}(n)$ denote  the smallest order (number of leaves, or vertices)
 of the maximum
agreement subtree of two  phylogenetic X-trees with $|X|=n$.
In 1992, Kubicka, Kubicki, and McMorris \cite{KKMcM} showed that
$c_1 (\log \log n)^{1/2} < {\rm mast}(n) < c_2 \log n$ with some
explicit constants.

The purpose of our note is to remove the squareroot sign from the lower bound.
This is achieved by changing the order of two combinatorial steps,
one resulting in taking logarithm twice, the other taking a squareroot.
Of course, the  squareroot sign after the $\log \log$ is no longer visible.

First we would like to exhibit a direct connection to Ramsey theory,
which might explain the large gap between the lower and upper bounds for
${\rm mast}(n)$. Let $R_2^k(n,\ell)$ denote the smallest integer $m$
such that for any coloration of the $k$-element subsets of any $m$-element
set with colors Red and Blue, there exists an $n$-element subset
of the $m$-element set, such that every   $k$-element subset
of the $m$-element set is colored Red, or there exists an $\ell$-element
subset
of the $m$-element set, such that every   $k$-element subset
of the $m$-element set is colored Blue (see Chapter 14 in \cite{lovasz}).

\noindent {\bf Claim.} ${\rm mast}[ R_2^4(n,6) ] \geq n$.\\
\noindent {\bf Proof.}
We first recall an observation from \cite{ban86} that for $|X|=6$, any two
 phylogenetic X-trees
share a quartet split. Given $T$ and $F$ arbitrary phylogenetic $X$---trees
 with
 $|X|=  R_2^4(n,6)$,
 color 4-subsets of $X$ Red, if they define the
same quartet split, otherwise Blue. No six elements of $X$ can have all
4-subsets Blue by the previous reference, so there are $n$ elements from X
 such that
all
their 4-subsets are colored Red. As the binary tree is determined by
its quartet splits, these $n$ elements span a size $n$ agreement subtree,
 thereby establishing the Claim.
\hfill $\Box$

This approach would give an explicit lower bound for ${\rm mast}(n)$ in the form
of a multiply-iterated logarithm, much weaker than
$c_1 (\log \log n)^{1/2}$.

Before proving our result, we quickly show $c_1 (\log \log n)^{1/2} < {\rm mast}
(n)$
following the approach in the 1992 paper by Kubicka, Kubicki, and McMorris \cite{KKMcM}.
Recall that a {\it caterpillar} is a tree, which has a path such that every
leaf has a neighbor on the path (for example, the tree $F$ in Fig. 1).
Let us
be given two phylogenetic X-trees $T$ and $F$ with $|X|=n$. As our trees
are binary, the diameter of $T$ is at least $c_3\log n$. Therefore $T$
must have an induced binary caterpillar subtree with leaf set $Y$,
such that $|Y|\geq c_3\log n$. Consider the induced binary subtree
$F|_Y$, which must have diameter $\geq c_4 \log  \log n$. Like we argued
before, there should be a $Z\subseteq Y$ such that $F|_Z$
is a caterpillar  and $|Z|\geq  c_4 \log  \log n$. Notice that
$T|_Z=(T|_Y)|_Z$ is also a caterpillar. Recall the Erd\H os-Szekeres Theorem
(Ex. 14.15 in \cite{lovasz}) for
 sequences: two sequences composed from the same $k^2+1$ items have either
a common $k+1$ length subsequence, or they have a
 common $k+1$ length subsequence after reversing the order in one sequence.
As caterpillar trees can be understood as sequences of their leaves,
two caterpillar trees with the same $k^2+1$ leaves contain
size $k+1$ agreement subtrees. Apply this with the largest $k$
such that $k^2+1 \leq c_4 \log  \log n$.

Before turning to our main result, we need some definitions. We say that
a phylogenetic X-tree $T$ is {\it drawn on the plane} if it is drawn as a
plane graph. The {\it circumference of phylogenetic tree drawn on the plane}
is the
cyclic permutation of  $X$, the leaf set,  as we walk around $T$ clockwise.
 This
 concept has been been a useful combinatorial tool elsewhere (see, for
 example,
\cite{semste}) and we illustrate it here in Fig. 1 by noting that
the circumference of this drawing of $T$ is the cyclic permutation
$(1,4,3,6,5,2)$.

Note that for $Y\subseteq X$  the  induced binary subtree of $T$  (by $Y$)
 has a
natural drawing following steps (a) and (b) by deleting edges and vertices
from the plane drawing, and then removing the vertex designation of
vertices of degree 2, but keeping the curve representing  the path for
representing the new edge.  For this natural drawing of $T|_Y$, the
circumference is the circumference of the drawing of $T$ restricted to $Y$.
 For
the tree $T$ in Fig. 1,
and the subset $Y=\{2,3,4,6\}$ the circumference of the induced drawing
of $T|_Y
$ is cyclic permutation $(2,4,3,6)$ (the same as the circumference of the given drawing of $G$)
 while the circumference of
the induced drawing of $F|_Y$ is the cyclic permutation $(2,3,4,6)$.

\begin{theorem}
\label{mainthm}
For a constant $c>0$, we have:
$$c \log \log n < {\rm mast}(n).$$
\end{theorem}
\noindent {\bf Proof.}  Take two arbitrary phylogenetic X-trees,
$T$ and $F$,  with
$|X|=n$
and draw them in the plane. Cut the resulting circumferences anywhere
to obtain two (linear) permutations of $X$. By the
Erd\H os-Szekeres Theorem, there is subset $U\subseteq X$, such that
the two permutations either put $U$ into the same linear order,
or into opposite linear order, and $|U|\geq c_5 n^{1/2}$.
Like in the proof explained before the theorem, $T|_U$ has diameter
$\geq c_3\log |U|\geq c_6\log n$. Therefore $T|_U$ has an induced
binary subtree which is caterpillar, with leaf set $V$, such that
$|V|\geq c_6\log n$. Consider now the induced binary subtree
$F|_V$. The diameter of  $F|_V$ is at least $c_3 \log |V|\geq c_7
\log\log n$, and therefore there should be a $Z\subseteq V$, such that
$F|_Z=(F|_V)|_Z$ is a caterpillar and $|Z|\geq c_7
\log\log n$. Both $T|_Z$ and $F|_Z$ are caterpillars. By the choice of
$U$, these two caterpillars have the same or mirror image  circumferences.
In the second case, starting this proof with the mirror image of the
drawing of $F$, we can make sure that the caterpillars  $T|_Z$ and $F|_Z$ have
identical circumferences. Taking the longest path from $T|_Z$ (resp. $F|_Z$),
this path partition the $|Z|-2$ non-endpoint leaves of $T|_Z$ (resp. $F|_Z$)
 into two classes,
corresponding to the two sides. We have two 2-partitions of $|Z|-4$
or more elements into two classes - it is easy to see that some partition
classes must have at least $(|Z|-4)/4$ elements in common, say $W$.
Now $T|_W=(T|_Z)|_W$ and $F|_W=(F|_Z)|_W$ are the common induced binary
subtree of $T$ and $F$, and $|W|\geq  c_8
\log\log n$.
\hfill $\Box$

\noindent {\bf Remark.}
It would be interesting to see whether Theorem~\ref{mainthm} can be
 tightened. In particular, it is conceivable that the much stronger bound
 $c' \log(n) < {\rm
mast}(n)$ holds, which would  be best possible, up to the constant factor.


\begin{thebibliography}{99}


\bibitem{ban86} Bandelt, H. -J. and Dress, A. W. M. (1986).
  Reconstructing the shape of a tree from observed dissimilarity data.
  {\em Advances in Applied Mathematics}, {\bf 7}: 309--343.

   \bibitem{colonius} Colonius, H. and Schulze, H. H. (1981). Tree structures
  for proximity data. {\em British Journal of Mathematical and
    Statistical Psychology}, {\bf 34}: 167--180.

\bibitem{fgordon}
Finden, C.R.  and  Gordon, A.D. (1985). Obtaining common pruned
trees, {\em J. Classification} {\bf 2}:  255-116.


\bibitem{GKKMcM}  Goddard, W. Kubicka, E., Kubicki, G. and McMorris, F.R.
 (1994). The agreement metric for labeled binary trees,
                 {\em Mathematical Biosciences} {\bf 123}: 215 - 226.



\bibitem {gordon} Gordon, A.D. (1986).  Consensus supetrees: the
synthesis of rooted trees containing overlapping sets of labelled leaves.
 {\em J. Classification} {\bf 3}: 335--348.



   \bibitem{KKMcM} Kubicka, E., Kubicki, G., and McMorris, F. R. (1992). On
  agreement subtrees of two binary trees. {\em Congressus Numerantium},
  {\bf 88}: 217--224.


\bibitem{KKMcM95} Kubicka, E., Kubicki, G. and McMorris, F.R. (1995).
 An algorithm
to find agreement subtrees, {\em J. Classification}  {\bf 12}: 91--99.



\bibitem{lovasz} Lov\'asz, L. (1993). {\em Combinatorial Problems
and Exercises}, 2nd ed. North-Holland.

\bibitem{semste} Semple, C. and Steel, M. (2004). Cyclic
permutations and evolutionary trees.
{\em Advances in Applied Mathematics} {\bf 32(4)}:  669-680.

\bibitem{stewar} Steel, M. and Warnow, T. (1993). Kaikoura tree theorems:
  computing the maximum agreement subtree. {\em Information Processing
    Letters} {\bf 48}: 77--82.


\bibitem{PAUP} Swofford, D. L. 2003. PAUP*. Phylogenetic Analysis
 Using Parsimony (*and Other Methods). Version 4. Sinauer Associates,
 Sunderland, Massachusetts.


\end{thebibliography}
\end{document}